# Democracy, Complexity, and Science

## Exploring Structural Sources of National Scientific Performance

*Accepted for Publication in Science and Public Policy*

Travis A. Whetsell[1], Ana-Maria Dimand[2], Koen Jonkers[3], Jeroen Baas[4], Caroline S. Wagner[5]

[1] Florida International University, School of International and Public Affairs, Department of Public Policy & Administration, Miami, FL, USA, Travis.whetsell@FIU.edu
[2] Boise State University, School of Public Service, Boise, ID, USA
[3] Knowledge for Growth, Finance and Innovation Unit, Joint Research Centre (JRC), Brussels, Belgium
[4] Elsevier B.V., International Center for the Study of Research, Radarweg 29, Amsterdam, The Netherlands
[5] The Ohio State University, John Glenn College of Public Affairs, Columbus, OH, USA

**Abstract:** Scholars have long hypothesized that democratic forms of government are more compatible with scientific advancement. However, empirical analysis testing the democracy-science compatibility hypothesis remains underdeveloped. This article explores the effect of democratic governance on scientific performance using panel data on 124 countries between 2007 and 2017. We find evidence supporting the democracy-science hypothesis. Further, using both internal and external measures of complexity, we estimate the effects of complexity as a moderating factor between the democracy-science connection. The results show differential main effects of economic complexity, globalization, and international collaboration on scientific performance, as well as significant interaction effects that moderate the effect of democracy on scientific performance. The findings show the significance of democratic governance and complex systems in national scientific performance.

**Highlights**
- Scholars suggest the hypothesis that democratic governance enhances the scientific performance of nations *vis a vis* value congruency and structural complexity.
- The rise of scientific systems in countries with authoritarian governance and the scientific decline of democratizing countries challenges the hypothesis.
- This study tests the hypothesis using data on democracy, complexity, and science on a panel of 124 countries over an eleven-year period.
- The results provide support for the hypothesis but suggest more nuanced moderating effects of complexity on the democracy-science relationship.

**Acknowledgements:** Early research presented at Sapienza University in Rome, Italy for the 17th International Conference on Scientometrics & Informetrics. Thank you to Mario Scharfbillig and Daniel Vertesy for comments on an earlier version of the manuscript.

**Disclaimer:** This article does not necessarily represent the official views of the European Commission. The European Commission nor anyone acting on its behalf can be held responsible for any use of the data or analysis contained herein.

**CRediT author statement:** Travis Whetsell: Writing, Conceptualization, Methodology, Data Curation, Formal analysis, Visualization. Ana-Maria Dimand: Writing, Data Curation, Formal analysis. Koen Jonkers: Writing, Conceptualization, Methodology. Jeroen Baas: Writing, Data Curation. Caroline Wagner: Writing, Conceptualization, Supervision

# 1. Introduction

Studies of science policy have long suggested that political, economic, and scientific systems are interdependent, but the nature of the relationship and the direction of causality remain unclear. Strasser (2009) suggested that society and culture are determinants of the construction of science. Sociologists and philosophers of science, such as Merton (1973), Popper (1966), and Kitcher (2003), asserted that democracy is conducive to the advancement of scientific research. Recent research on international scientific openness and mobility also suggest a positive effect on national systems of science (Wagner et al 2018a, 2018b). Weisner et al. (2018) suggest democracies may permit greater structural complexity and stability, creating conditions wherein scientific inquiry can thrive. If it true that science operates as a modular loosely coupled system with an emergent networked architecture (e.g. Simon 1966), then the mutual complexity of democracy and science may produce benefits for both. However, empirical studies have yet to adequately address questions about the relationship between forms of government and the performance of national systems of science.

At the macro-level, the relationship between democracy and science has mostly been analysed conceptually by Western scholars often with the use of exemplar cases during the Cold War. Parsons (1951), Ben-Joseph (1971) and Giddens (1987) suggested that science is integrated with the social structures and cultural traditions of society. Barber (1953) and Ferris (2010) reiterated the Enlightenment view that liberal societies and science share common values. Thus, democracies often promote economic growth through scientific progress. Merton (1978) and Harrison & Johnson (2009) argued that both science and democratic societies share the unique feature of universalism, while autocratic societies undermine this principle through the exaltation of political ideology. Popper (1966) argued that the open society permits greater contact with new ideas from abroad and those that arise internally.

While research on national innovation systems emerged to explain how governments, universities, and firms produce innovative outcomes during the last quarter of the 20$^{th}$ century (Freeman, 1987; Nelson & Rosenberg 1993; Lundvall et al., 2002; Godin, 2009; Taylor 2016), others have increasingly pointed to the interconnectedness of national level institutions forming a complex global system of science and technology (Tyson, 1988; Archibugi et al., 1999; Carlson, 2006; Wagner 2009). Recent research suggests international scientific openness and mobility have positive effects on country level scientific performance (Wagner et al. 2018a, 2018b). However, the relationship between varieties of democracy and science and technology remains unclear (Gao et al. 2017).

We find arguments about the necessary relationship between democracy and science compelling, yet the coexistence of successful scientific systems coupled with authoritarian governance, e.g. the Soviet Union or contemporary China, challenges the hypothesis. This article builds on theory in science and technology policy, leveraging national measures of scientific performance to measures of democratic governance. Theory suggests that democratic governance may enhance system complexity (Weisner et al. 2018), which we operationalise using measures of economic complexity, globalization, and international collaboration intensity to test the democracy-science compatibility thesis. While previous research has focused on national openness and mobility (e.g. Sugimoto et al, 2017; Wagner et al 2018a, 2018b), the current article conceptualizes these concepts within the broader category of *external* complexity, treating economic complexity as a measure of *internal* complexity. By emphasizing the complex interdependence of national systems and the larger global system, this article builds a unified theoretical approach to the study of scientific performance.

This article tests two empirical research questions using fixed effects regression on a large panel of 124 countries over an eleven-year period: 1) what is the relationship between democratic governance and the scientific performance of nations in the international system;

and 2) how do internal and external complexity moderate the democracy-science relationship. To address these questions, this article combines longitudinal data from the Varieties of Democracy (V-Dem) Project with bibliometric data on scientific performance from Elsevier/Scopus, as well as data from the MIT Observatory of Economic Complexity, the KOF Swiss Economic Institute, and the World Bank.

**2. Literature**

    **2.1**. *Governance & Science*. Parsons (1951) suggested that the development of science is integrated with the social structures and cultural traditions of a given society. In this sense, there is a suggested compatibility between the social structures of science and society. Structural determinants of science *vis-à-vis* society are particularly salient regarding the institutionalization of science in organizations and the conceptualization of the "scientist" in terms of a formal occupational role.

    This article emphasizes the broader mechanisms outside of science, e.g. national system of governance, economic structure, cultural values, etc. This article focuses on science in society, and not the society of science. Variance in socio-cultural structure within which science operates has direct implications for the structure and performance of science. Building on Parsons, Barber (1953) suggests that the cultural values of liberal society are congruent with those of science. The cultural values that Barber identifies as congruent with science are rationality, utilitarianism, universalism, individualism, and progress. Thus, the congruency between democracy and science on structure and values produce more favourable conditions for the development and flourishing of science.

    Merton (1973) emphasizes universalism in the value congruency between democratic forms of government and the enterprise of science, where universalism entails that "truth claims, whatever their source, are to be subjected to *pre-established impersonal criteria*:

consonant with observation and with previously confirmed knowledge" (Merton, 1973, p. 270, emphasis in original). For Merton, the universalism of science is compatible with democracy but clashes with totalitarianism based on political, religious, or ethnic identity. Under totalitarian regimes, science becomes the "handmaiden" of an ethic that subordinates scientific theory to political identity (Merton 1938). In contrast to more decentralized democratic governments, Merton (1973) suggests totalitarian governments impose centralized control that infringes upon free inquiry: "In modern totalitarian society, anti-rationalism and the centralization of institutional control both serve to limit the scope provided for scientific activity" (Merton 1973, p. 278). In this sense, democracy decentralizes power to individual citizens, rather than centralizing it in the state. Decentralized structure permits free association and expression based on universalistic truth claims that are not subordinated by the privileged particularisms of whatever dominant political ideology is ascendant.

Like the Mertonian argument for the compatibility of democracy and science, Popper (1966) suggested that closed autocratic societies tend to be philosophically situated on fixed but fragile historicist visions. Unconstrained scientific progress threatens to undermine these fragile visions and so poses a significant threat to the centralized power of autocratic regimes. As Popper suggested, antiquated scientific conceptions often emphasized "methodological essentialism" that attempts to identify unchanging Platonic forms underlying empirical phenomena. While compatible with an autocratic Republic envisioned by Plato, contemporary science challenges these fixed historicist visions. Thus, scientists must be free to pursue 'truth,' (or validity) and open societies provide a better context for the questioning and experimentation of scientific activity. Autocratic regimes constrain the activities of scientists to those that do not challenge centralized social control, while democratic regimes permit greater autonomy in the pursuit of scientific truth. According to Mokyr (2017) the variation, openness, and competition between ideas that characterised Western European societies can explain why the scientific and technological revolution occurred there rather than in China.

The democracy-science compatibility thesis, however, is not a given among scholars of science and technology studies. Stephan (2012) softens this view by pointing out that democratic government seeks new knowledge for socially beneficial outcomes, but without specifying outcomes beforehand. Democracy may also pose unique threats to scientific advancement when public opinion shifts against topics of inquiry, such as the theory of evolution (Kitcher 2003).[1]

Gao et al. (2017) recently attempted to test a hypothesis, orthogonal to our own, of whether democracy is conducive to technological innovation. They were unable to show such a relationship between democracy measures and patent data in their analysis. They do however confirm an emerging body of literature that suggests openness can be a strong factor in technological progress, as noted above. Indeed, national openness appears to be associated with scientific impact, where *openness* is operationalized through international co-authorship data and mobility statistics on the scientific workforce (Wagner et al. 2018a, 2018b; Chinchilla-Rodríguez 2018; Robinson-Garcia et al. 2019). However, we remain sceptical that Gao et al.'s (2017) finding generalizes to a non-significant relation between democratic governance and science. Following on the above conceptual work and research empirical work, we advance a positive hypothesis regarding the theoretical relationship between democracy and science.

*Hypothesis 1: Democratic governance is positively associated with the scientific performance of countries in the international system.*

---

[1] Nahuis and Van Lente (2008) suggest the direction of causality might be reversed, where science may undermine or displace democracy. Studies of science and technology have examined public participation in science (Lengwiler 2008), responsibilities of science given a democratic context (Stilgoe, Owen, & Macnagthen, 2013), and the potential dangers of science in undermining democracy (Durant, 2011). This is a sensible concern given the potential dangers that science poses to human subjects (Proctor 1988), as well as the power dynamics associated with the elevation of expert over the popular will (Brown 2009).

Much of the conceptual and comparative casework on the democracy-science hypothesis outlined above suggests that specific values or structures may be implicated to a greater degree than other less salient points of contact. The question is, what specific values or structures of society are compatible with science? Among the values that recur in this literature are openness, objectivity, autonomy, and universality. Further, decentralization of power in socio-political structures is also compatible with the decentralization of scientific activity in myriad disciplines and fields, as well as the system of peer-review that decentralizes academic authority to the scientists. Each of these elements have analogues in various conceptualizations of democracy.

Numerous indicators have been proposed to measure forms of democracy, e.g. Polity IV and the Economist Democracy Index. Recently, the Varieties of Democracy (V-Dem) project has provided detailed longitudinal data on several high-level indexes of democracy aggregated from hundreds of empirical measures. The overarching concept of electoral democracy in the V-Dem data is based on Dahl's (1998) notion of polyarchy, or "rule by many". The V-Dem project defines democracy in terms of the following five elements: "(1) elected officials; (2) free, fair and frequent elections; (3) freedom of expression and alternative sources of information; (4) associational autonomy; and (5) inclusive citizenship (universal suffrage)" (Coppedge et al. 2016: 582). Under this conceptual scheme, the electoral democracy index is the basic concept of polyarchy, which is additive in the sense that the other four indexes also include this index. This study operationalizes the first hypothesis using the democracy indexes of the V-Dem project.

Quantification of scientific performance at the country level is a relatively new advancement that has traditionally been limited to measures of the inputs of science, such as national spending, number of researchers and institutions, etc. However, developments in bibliometrics over the past few decades has enabled the measurement of outputs through publication meta-data. This study assesses the variation in output quality of national scientific systems using more recent advances in bibliometrics, building on approaches that measure scientific impact by publication citations (Wagner et al. 2018). To account for different citations practices between years and fields, a normalization technique is applied, resulting in a Field-Weighted Citation Impact (Purkayastha et al. 2019). Individual citation scores per article are fractionalized and in national aggregations weighted based on those fractional shares of authorship (Perianes-Rodriguez, Waltman & van Eck, 2016). This so-called fractional counting addresses the relative higher representation that full counting would have on publications that are co-authored by many authors from different origin: it results in an average for national aggregates, in line with the global average.

**2.2.** *Internal Complexity & Science.* The argument for the democracy-science connection may be extended through an analysis of structural complexity, where the organizational/institutional structure of science is critical to progress. The contemporary structure of science is decentralized in a modular global network with self-similar structure up and down levels of analysis (Wagner & Leydesdorff, 2005; Wagner, Whetsell, & Leydesdorff, 2017). As Simon (1996) observed, modular, loosely coupled hierarchies can be more efficient at handling uncertainty than fully centralized hierarchies. Kontopoulos (2006) refers to this type of structure as a heterarchy. A core activity of science is to create knowledge at the margins of uncertainty. Therefore, decentralized modular institutional systems are perhaps better suited to the flourishing of science than centralized hierarchies. Wiesner et al. (2018) suggest that democracies may permit greater structural complexity and stability. Since science operates as a complex, loosely coupled system with an emergent networked architecture (Wagner, 2008; Simon 1996), such systems may permit greater internal complexity in its cultural, social, and economic systems.

Research on economic complexity suggests that national capacity for the integration of diverse sources of knowledge is a unique resource explaining differences in gross domestic product (GDP) between nations (Hidalgo and Hausman 2009). National capacity for economic complexity may also contribute to relative differences in national systems of science. The Economic Complexity Index (ECI) is defined as the knowledge intensity of a country's export products. ECI+ modifies the variable by accounting for the difficulty of exporting products for the country. While the ECI is based on an analysis of the knowledge intensity of the export products in a nation's economy, this measure may serve as a useful proxy for the availability and complexity of knowledge in a country in general. As Nelson and Rosenberg (1993) suggested, science both leads and follows technology, i.e. the two are closely intertwined. Hence, the complexity of a country's technological products may be partially indicative of the complexity of the institutional structure of scientific production.

This article conceptualizes economic complexity as a form of internal complexity, which signals the level of internal differentiation and diversity of a country's research institutions. The decentralization of political power in democracies may be more compatible with the internal structural complexity necessary for scientific performance. As such, economic complexity may enhance the democracy-science relationship by enriching systems of science with dense flows of information and greater serendipity in the generation of novel combinations of existing knowledge (e.g. Wagner, Whetsell, and Mukherjee 2019). This study advances the following hypothesis.

> *Hypothesis 2: Economic complexity moderates the democracy-science relationship among countries in the international system.*

**2.3.** *External Complexity & Science.* The degree to which a society creates new knowledge does not only depend on the extent to which indigenous flowers can bloom. New ideas, technologies and values may also originate from beyond the national context. Contacts with other cultures through trade and cultural exchange have been at the basis for many scientific and technological advances over the course of the past centuries, and the degree of receptiveness to foreign ideas may explain a large share of the difference in long term scientific and technological performance between countries over time (Mokyr, 2017). Taylor (2016) argued that the extent to which societies are integrated in the global system measured in terms of trade, investments, culture and political ties following the KOF Globalisation index (Dreher, 2006; Savina et al., 2018) can explain a relatively large degree of the variation in technological performance of countries. This article conceptualizes globalization as a form of external complexity. Just as decentralization in political authority may be compatible with decentralization, differentiation, and diversification of scientific institutions, so too it may be compatible with the proliferation of institutional linkages that span national boundaries.

There are many ways to conceptualize the relative globalization of a country. Traditional realist international relations theory may conceptualize globalization as an extension of state power (Mearsheimer 2001), while liberal internationalist theory might view it as a set of international regimes designed to reduce the transaction costs of cooperation (Keohane 1984). In keeping with the theme of complexity, the globalization of a country may be conceptualized as a type of external complexity, where the internal knowledge intensity of an economy is integrated with a macro-system that is in some sense separate or above the national systems. In this sense, we may conceptualize globalization as an emergent property of the interaction between countries.

While the hypothesis has never been tested, an extension of existing theory suggests globalization might play a moderating role in the democracy-science relationship. Democracies might fare better in their pursuit of science, not only because of their internal culture and complexity, but also through increased contact with international sources of knowledge and their integration with the global network of scientific activity. As such, we advance the following hypothesis which regards globalization as a dynamic that integrates national systems with the broader international system.

> *Hypothesis 3: Globalization moderates the democracy-science relationship among countries in the international system.*

Another type of external complexity that has received a great deal of attention in recent decades is the ever-increasing rate of international collaboration in scientific research. Numerous studies have documented the rise of internationally co-published research (e.g. Wagner and Leydesdorff, 2005; Wagner, Whetsell, Leydesforff, 2017). Studies have also documented the increasing impact on international collaboration on scientific outcomes such as the influence and novelty of research (Sugimoto et al, 2017; Chinchilla-Rodríguez 2018; Robinson-Garcia et al. 2019; Wagner, et al. 2018a, 2018b; Wagner, Whetsell, and Mukherjee 2019). We extend theory on international collaboration by considering it as a type of globalization that is specific to the sciences. While the KOF globalization index is a composite of numerous social, economic and political indicators, international research collaboration is more specifically bounded as the co-publication of scientific papers of authors with distinct national affiliations. As such, we expect that democratic countries tend to be more open and collaborative with other countries in the international system, and that such increases in external complexity may be beneficial for the greater flourishing of scientific performance. Thus, we hypothesize that the relationship between democracy and science is enhanced by international collaboration.

> *Hypothesis 4: International collaboration moderates the democracy-science relationship among countries in the international system.*

# 3. Methods

**3.1.** *Data:* The data for this research comes from multiple sources. For primary measures of democracy, we use the Varieties of Democracy (V-Dem) panel data (Coppedge et al. 2019) which were available 2008 to 2017 on 162 countries. For measures of globalization, we include the KOF Swiss Economics Institute's Globalization Index (Dreher, 2006; Gygli et al., 2019). For economic complexity, we use the Economic Complexity Index (ECI+) from the MIT Observatory for Economic Complexity (Hidalgo and Hausmann 2009; Simoes and Hidalgo 2011). We also use data from World Bank Open Data for control variables, including GDP per capita and population size. For measures of scientific production and impact we use Elsevier/Scopus bibliometric data aggregated at the national level, which was previously employed by Wagner et al. (2018b).

**3.2.** *Variables:* The dependent variable in all regression models is national scientific performance, which is operationalized using Elsevier's fractional field weighted citation index (Frac_FWCI). FWCI is a recently developed indicator of citation impact that accounts for variability across scientific fields by dividing observed citation counts by expected counts for the field (*see* Purkayastha et al. 2019). Fractional (Frac) counting means that proportions of authors are summed when calculating volumes and are weighed when relative citation impact is aggregated to the total citation impact of the country (Waltman & van Eck 2015; Wagner et al. 2018a, 2018b). More recently, a modified fractional counting (MFC) approach was introduced that addresses variation between different co-authorship practices between fields (Sivertsen, Rousseau & Zhang, 2019). The MFC approach is evaluated for the use in counting contributions of entities but has not yet been applied to citation indicators and is therefore not yet applied in our methodology but may provide an enhancement in the future. A brief description of all variables is presented in Table 1.

For the key independent variables that capture national democratic governance, this article uses the Varieties of Democracy Project (V-Dem Project) data, which include five high level indices of democracy which are aggregated from mid-level and lower-level measures created from over 120 variables (Coppedge et al. 2019). The high-level indictors used in this article are polyarchy, liberal, participatory, deliberative, and egalitarian democracy indexes. Polyarchy is the basic indicator of electoral democracy adapted from Dahl (1998). The V-Dem Project (2019:39) defines democracy in the following manner.

> *"The electoral principle of democracy seeks to embody the core value of making rulers responsive to citizens, achieved through electoral competition for the electorate's approval under circumstances when suffrage is extensive; political and civil society organizations can operate freely; elections are clean and not marred by fraud or systematic irregularities; and elections affect the composition of the chief executive of the country. In between elections, there is freedom of expression and an independent media capable of presenting alternative views on matters of political relevance." (V-Dem Project,* 2019:39*)*

For moderating variables, we include a measure of internal complexity operationalized using the Economic Complexity Index (ECI+) measure from MIT's Observatory on Economic complexity.[2] The ECI measures the relative intensity of knowledge at the level of the country's economy by assessing the complexity of a country's export products (Hidalgo and Hausmann 2009). ECI+ modifies the variable by accounting for the difficulty of exporting for the country (Albeaik et al. 2017).[3] For moderating variables of external complexity, we use two measures: globalization and international collaboration intensity. The first is the KOF Globalization Index, a panel data set, which is a high-level aggregation of numerous empirical measures across political, economic, and social globalization (Dreher, 2006; Gygli et al., 2019). To measure international collaboration intensity, we divide the number of international papers a country co-produced in a given year by the total number of papers for the country-year, producing a percentage of the total country papers which are internationally co-published.

---

[2] Albeaik, et al. (2017:3-4) define ECI+ as such: "the complexity of an economy as the total exports of a country corrected by how difficult it is to export each product and by the size of that country's export economy".
[3] Alternative measures of economic complexity have also been proposed, see Tacchella et al. (2012).

**Table 1: Summary Statistics**

| Variable | Source | Brief Description | Years | Obs | Mean | Std. Dev. | Min | Max |
|---|---|---|---|---|---|---|---|---|
| (1) Polyarchy | V-Dem Project | Rule by many elected officials | 2007-17 | 1,364 | 0.585 | 0.267 | 0.016 | 0.948 |
| (2) Liberal Democracy | V-Dem Project | Individual and civil rights | 2007-17 | 1,364 | 0.478 | 0.277 | 0.031 | 0.914 |
| (3) Participatory Democracy | V-Dem Project | Suffrage and participation | 2007-17 | 1,364 | 0.388 | 0.217 | 0.011 | 0.832 |
| (4) Deliberative Democracy | V-Dem Project | Mode of political decision making | 2007-17 | 1,364 | 0.485 | 0.264 | 0.016 | 0.912 |
| (5) Egalitarian Democracy | V-Dem Project | Distribution of rights, resources, and power | 2007-17 | 1,364 | 0.469 | 0.251 | 0.033 | 0.9 |
| (6) Economic Complexity | MIT Economic Observatory | Knowledge intensity of products | 2007-15 | 844 | 0.232 | 0.908 | -3.532 | 1.747 |
| (7) Globalization Index | KOF Swiss Econ. Institute | Integration with international system | 2007-17 | 1,237 | 68.689 | 13.072 | 33.226 | 91.313 |
| (8) Collaboration Intensity | Elsevier/Scopus | Percent of papers that are internationally co-published | 2007-18 | 1,404 | 0.529 | 0.171 | 0.142 | 0.957 |
| (9) Total Publications | Elsevier/Scopus | Total publications by fractional count | 2007-18 | 1,404 | 19163.1 | 57215.97 | 88.98 | 529524.5 |
| (10) GDP Per Capita | World Bank | GDP divided by population size | 2007-17 | 1,214 | 24751.5 | 22982.85 | 807.44 | 140037.1 |
| (11) Population Size | World Bank | Population size | 2007-17 | 1,238 | 5.78E+07 | 1.76E+08 | 3.12E+05 | 1.39E+09 |
| (12) Frac_FWCI | Elsevier/Scopus | Fractional field weighted citation index | 2007-18 | 1,404 | 0.79 | 0.312 | 0.154 | 1.636 |

For control variables, we include the fractional count of the number of papers produced by the country, which is a measure of the publication volume of the country and a measure of the size of the scientific system of the country. The fractional count accounts for the portion of the publications that are not inflated by international collaboration. We also include population size, and per capita GDP – log transformations of these variables were also tested but did not result in significant effects for population or GDP.

**3.3.** *Sample*: One limitation of Frac_FWCI aggregated at the national level, is that it is susceptible to outliers, where very low productivity countries co-publish all/most of their publications with high productivity nations. Thus, many otherwise unproductive nations appear to have very high impact numbers due solely to their collaboration with other productive nations. To disentangle part of this bias on Frac_FWCI we filtered out observations by analysing the distribution of full publication numbers of all countries, which shows a skewed distribution with a mean value of 12252.23, with the 25th percentile at 31 papers, the 50th percentile at 279.5, and the 75th percentile at 3223. All subsequent analysis is performed on observations equal to or above 279.5 publications. This reduced the sample size, when including all model variables, from 164 to 124 countries. Further, filtering out low productivity observations produced a less skewed distribution in the dependent variable (Frac_FWCI). The difference in frequency distributions before and after removing outliers based on total publications is shown in the appendix.

**3.4.** *Methods:* Because the data provided repeated observations on each country over a multiyear time period, the study was able to leverage panel data techniques to estimate relationships between the variables. The data coverage differs by variable; therefore the data set is classified as unbalanced panel data. To accommodate the data structure, to account for fixed country and year effects, and to estimate the within-unit effect of a change in democracy on scientific performance, we chose to utilize a two-way fixed effects model. The fixed effects model is an econometric model commonly used to analyse panel data, which Wooldridge (2010, 285) refers to as a "unobserved effects model", represented in the equation below.

$$y_{it} = x_{it}\beta + c_i + u_{it}, \ t = 1,2,\ldots,T,$$

The model estimates the effects on the dependent variable, $y_{it}$ (where $i$ is the country cross-section and $t$ is the year), of observed variables of interest, represented by $x_{it}\beta$ and the error term represented by $u_{it}$. However, the model also adds the parameter $c_i$ which includes an estimate for each country $i$ in the model. This controls for any effect on the dependent variable of unobserved time-invariant factors that are unique to the country. In the present context, parameter estimates in a fixed effects model are interpreted as *within* country effects which ignore cross-sectional effects *between* countries. Contrary to a model which analyses cross-sectional correlations *between* countries, where the question is whether more or less democratic countries have higher or lower scientific performance on average, in the fixed effects model the relevant question is whether an increase or decrease in democracy is associated with a commensurate change in its scientific performance over time *within* the country. The fixed effects model is referred to as "two-way" when both the cross-sectional parameter (country) is included and the time (year) parameter is included. Further, the model also uses cluster robust standard errors to account for heteroskedasticity and serial correlation. However, reverse causality and omitted time-varying variable bias remain threats to internal validity for observational studies. Additional robustness checks were conducted using generalized method of moments analysis (GMM) and Bayesian mixed models, presented in the Appendix.

## 4. Results

The analysis begins with pairwise correlations between all variables used in the analysis. Because the data include repeated observations for each country over the time period, the variables were collapsed on the country mean for the full time period before pair-wise correlations were conducted. The correlations in Table 2 show moderate to strong correlations between all five democracy indexes and scientific performance (Frac_FWCI). The correlations also show moderate to strong correlations between economic complexity and globalization on Frac_FWCI with no significant association for collaboration intensity. Further, the correlations also show moderate to strong correlations between the democracy indexes, economic complexity, and globalization, with a negative but insignificant correlation with collaboration intensity. The democracy scores are also strongly correlated with each other, suggesting that countries scoring higher on one dimension are also likely to score higher on others while also indicating differential effects of each on scientific performance. The results of the pairwise correlations warrant further analysis.

**Table 2 – Pairwise Correlations**

| Variables | (1) | (2) | (3) | (4) | (5) | (6) | (7) | (8) | (9) | (10) | (11) |
|---|---|---|---|---|---|---|---|---|---|---|---|
| (1) Polyarchy | 1 | | | | | | | | | | |
| (2) Liberal_Dem | 0.980* | 1 | | | | | | | | | |
| (3) Participatory_Dem | 0.976* | 0.972* | 1 | | | | | | | | |
| (4) Deliberative_Dem | 0.976* | 0.988* | 0.963* | 1 | | | | | | | |
| (5) Egalitarian_Dem | 0.957* | 0.976* | 0.952* | 0.967* | 1 | | | | | | |
| (6) Econ_Complexity | 0.424* | 0.495* | 0.462* | 0.473* | 0.521* | 1 | | | | | |
| (7) Globalization_Ind | 0.643* | 0.713* | 0.663* | 0.688* | 0.749* | 0.799* | 1 | | | | |
| (8) Collab_Intensity | -0.135 | -0.151 | -0.143 | -0.125 | -0.149 | -0.511* | -0.357* | 1 | | | |
| (9) Tot_Publications | 0.126 | 0.164 | 0.142 | 0.171 | 0.152 | 0.347* | 0.212* | -0.382* | 1 | | |
| (10) GDP_Per_Capita | 0.218* | 0.324* | 0.232* | 0.311* | 0.375* | 0.490* | 0.528* | -0.073 | 0.13 | 1 | |
| (11) Population_Size | -0.097 | -0.091 | -0.088 | -0.059 | -0.125 | 0.156 | -0.072 | -0.389* | 0.634* | -0.122 | 1 |
| (12) Frac_FWCI | 0.446* | 0.528* | 0.463* | 0.536* | 0.528* | 0.457* | 0.508* | 0.173 | 0.260* | 0.572* | -0.04 |

Table Notes: * $p<0.05$; The table shows the pairwise correlations between all variables; the mean value by country over the full time period was used for each variable to account for within subject correlation; Since the collapsed (mean) 50th percentile for full publications is different than the non-collapsed data, we first filtered out observations below or equal to the 50th percentile, which is 207.25 publications (see sample section for explanation).

To visualize the correlation between our primary measure of democracy (polyarchy) and scientific performance, Figure 1 presents a scatterplot with a fitted line that shows the correlation from Table 2 of 0.446.

**Figure 1 – Scatterplot of Polyarchy and Scientific Performance**

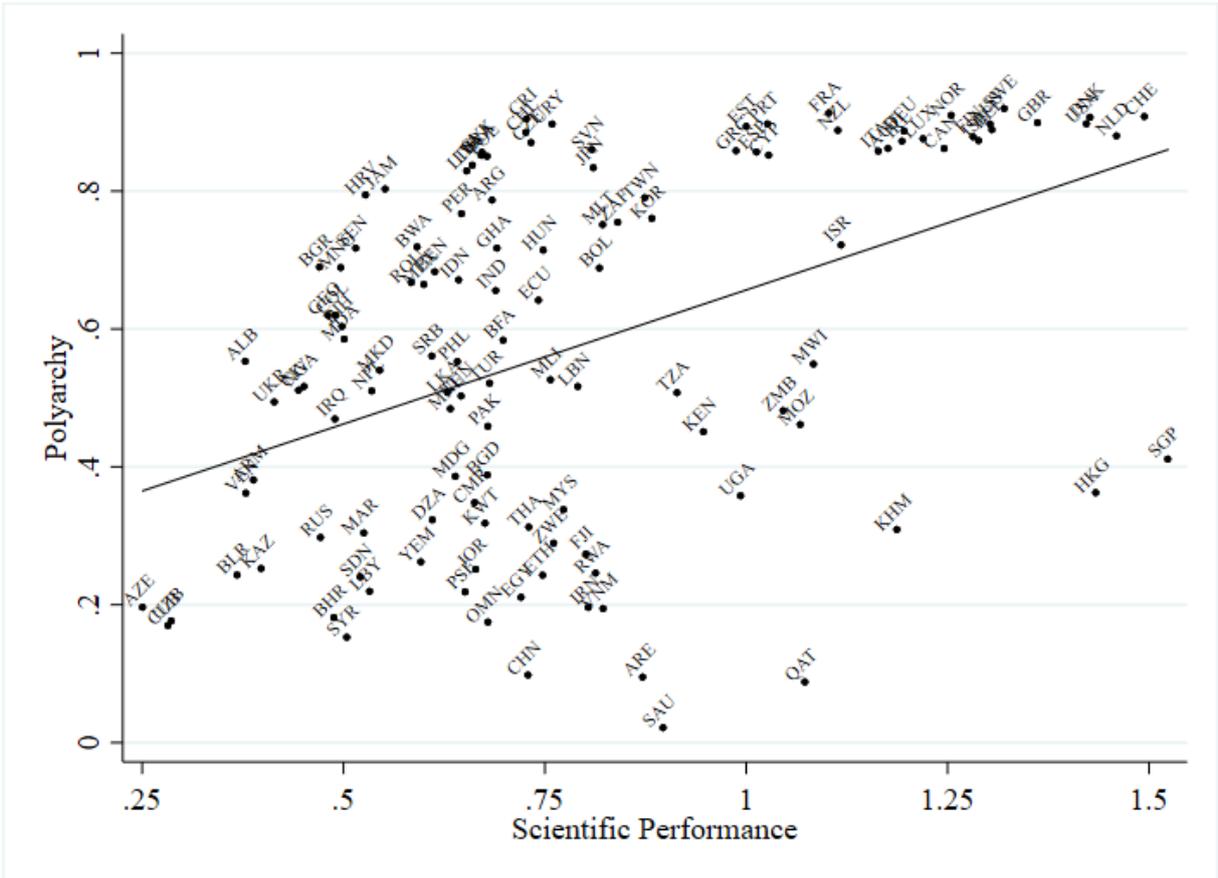

Figure Notes – Figure visualizes the correlation between polyarchy and scientific performance corresponding to the correlation in Table 2.

Next, the analysis leverages the panel structure of the data which allows modelling of effects over time. Table 3 shows the results of the two-way fixed effects models, where country and year are both treated as fixed factors in the analysis, and cluster robust standard errors are used to account for heteroskedasticity and serial correlation. Supporting H1, the results show consistent positive and significant effects for each democracy index on scientific performance across Models 1-5. Model 5 shows that Egalitarian Democracy has the largest estimate on scientific performance. The total publications (scientific volume) by the country also showed a positive and significant estimate on scientific performance. These results provide support for the first hypothesis concerning the democracy-science compatibility thesis.

**Table 3 – Effects of Democracy on Scientific Performance (2007-2017)**

|  | Model 1 | Model 2 | Model 3 | Model 4 | Model 5 |
|---|---|---|---|---|---|
| **Polyarchy** | 0.168** | | | | |
|  | (3.26) | | | | |
| **Liberal_Democracy** | | 0.157** | | | |
|  | | (2.82) | | | |
| **Participatory_Democracy** | | | 0.217** | | |
|  | | | (2.82) | | |
| **Deliberative_Democracy** | | | | 0.153** | |
|  | | | | (2.96) | |
| **Egalitarian_Democracy** | | | | | 0.267** |
|  | | | | | (3.36) |
| **Tot_Publications** | 0.000** | 0.000** | 0.000** | 0.000** | 0.000** |
|  | (3.11) | (3.10) | (3.13) | (3.12) | (3.14) |
| **GDP_Per_Capita** | 0.000 | 0.000 | 0.000 | 0.000 | 0.000 |
|  | (0.49) | (0.48) | (0.51) | (0.45) | (0.54) |
| **Population_Size** | -0.000 | -0.000 | -0.000 | -0.000 | -0.000 |
|  | (-0.34) | (-0.39) | (-0.33) | (-0.21) | (-0.36) |
| **Constant** | 0.637*** | 0.662*** | 0.649*** | 0.658*** | 0.606*** |
|  | (8.08) | (8.61) | (8.26) | (8.45) | (7.05) |
| **N** | 1188 | 1188 | 1188 | 1188 | 1188 |
| **#Countries** | 124 | 124 | 124 | 124 | 124 |
| **r2** | 0.116 | 0.113 | 0.114 | 0.116 | 0.119 |

Table Notes: The dependent variable is Frac_FWCI for all models. All models are two-way (country,year) fixed-effects models with country clustered standard errors; t statistics in parentheses; * $p<0.05$, **, $p<0.01$, *** $p<0.001$. GDP per capital and population size were also log transformed but did not produce significant effects. These models are subjected to robustness checks in the Appendix using GMM and Bayesian mixed effects.

Next, the analysis tests the moderation effects of the complexity indicators, including one measure of internal complexity (economic complexity) and two measures of external complexity (globalization and collaboration intensity). Variables have been standardized with mean zero and standard deviation one, as is common practice for interaction terms. Model 1 adds economic complexity to the basic polyarchy model, where both have positive estimates, but neither are significant. Model 2 interacts polyarchy with economic complexity, where the interaction term shows a positive and significant estimate and the main effect of economic complexity becomes significant. This result supports H2, which specifies that economic complexity enhances the democracy-science relationship. Model 3 adds the globalization index to the basic polyarchy model, where the main effect of polyarchy is positive and significant, while globalization is positive but insignificant. Model 4 adds the interaction term between polyarchy and globalization, showing a negative but insignificant estimate on the term. This result fails to provide support for the moderation effect of globalization on the democracy-science relationship, specified in H3. Model 5 adds collaboration intensity to the basic polyarchy model, where both variables show a positive and significant effect. Model 6 adds the interaction term between polyarchy and collaboration intensity, showing a negative and significant effect of the interaction on scientific performance. Again, this result contradicts H4, which specifies a positive moderation effect of collaboration intensity on the democracy-science relationship.

**Table 4 – Moderation Effects of Complexity (2007-2017)**

|  | Model 1 | Model 2 | Model 3 | Model 4 | Model 5 | Model 6 | Model 7 |
|---|---|---|---|---|---|---|---|
| **Polyarchy** | 0.025 | 0.009 | 0.042** | 0.046** | 0.032** | -0.043* | -0.039 |
|  | (1.60) | (0.48) | (3.18) | (3.05) | (2.75) | (-2.38) | (-1.98) |
| **Econ_Complexity** | 0.033 | 0.050* |  |  |  |  | 0.048* |
|  | (1.30) | (2.17) |  |  |  |  | (2.15) |
| **Polyarchy X Econ_Complexity** |  | 0.064* |  |  |  |  | 0.047* |
|  |  | (2.61) |  |  |  |  | (2.18) |
| **Globalization_Index** |  |  | 0.031 | 0.023 |  |  | 0.010 |
|  |  |  | (0.50) | (0.38) |  |  | (0.21) |
| **Polyarchy X Globalization_Index** |  |  |  | -0.024 |  |  | -0.063* |
|  |  |  |  | (-0.85) |  |  | (-2.06) |
| **Collab_Intensity** |  |  |  |  | 0.118** | 0.086** | 0.060* |
|  |  |  |  |  | (3.03) | (3.14) | (2.37) |
| **Polyarchy X Collab_Intensity** |  |  |  |  |  | -0.089*** | -0.074*** |
|  |  |  |  |  |  | (-5.14) | (-4.70) |
| **Tot_Publications** | 0.000** | 0.000** | 0.000** | 0.000** | 0.000** | 0.000** | 0.000** |
|  | (2.85) | (3.32) | (3.09) | (3.05) | (3.08) | (2.85) | (3.28) |
| **GDP_Per_Capita** | 0.000 | 0.000 | 0.000 | 0.000 | 0.000 | 0.000 | 0.000* |
|  | (1.45) | (1.63) | (0.58) | (0.60) | (0.40) | (1.14) | (2.13) |
| **Population_Size** | -0.000* | -0.000* | -0.000 | -0.000 | 0.000 | -0.000 | -0.000* |
|  | (-2.49) | (-2.52) | (-0.27) | (-0.20) | (0.13) | (-0.50) | (-2.03) |
| **Constant** | 0.710*** | 0.674*** | 0.708*** | 0.725*** | 0.817*** | 0.775*** | 0.795*** |
|  | (9.70) | (8.92) | (8.44) | (9.36) | (11.58) | (14.39) | (12.51) |
| N | 831 | 831 | 1167 | 1167 | 1188 | 1188 | 822 |
| #Countries | 100 | 100 | 121 | 121 | 124 | 124 | 99 |
| r2 | 0.119 | 0.150 | 0.113 | 0.116 | 0.191 | 0.284 | 0.323 |

Table Notes: The dependent variable is Frac_FWCI for all models. All models are two-way (country,year) fixed-effects with country clustered standard errors; Variables in interaction terms are standardized; t statistics in parentheses; N varies across models due to uneven data coverage of variables; * p<0.05, **p<0.01, *** p<0.001. These models are subjected to robustness checks in the Appendix using GMM and Bayesian mixed effects.

Model 7 includes all variables at once, showing a positive and significant interaction between polyarchy and economic complexity, and negative and significant interactions between polyarchy and globalization and collaboration intensity. This result provides support for H2 but contradicts H3 and H4. The results provide support for the general research question that complexity moderates the democracy-science effect.

To illustrate the interaction effects between polyarchy, economic complexity, globalization, and international percent, the analysis shows three margin plots in Figure 2, 3, and 4, corresponding respectively to Models 2, 4, and 6 in table 4. Figure 1 shows that the highest margin of democracy has the steepest positive slope on Frac_FWCI as the level of economic complexity increases. This indicates that as complexity increases the association between democracy and scientific performance is greater. As the level of polyarchy declines the association between economic complexity and scientific performance appears to reverse.

**Figure 2 – Interaction Effect of Polyarchy and Economic Complexity on Scientific Performance**

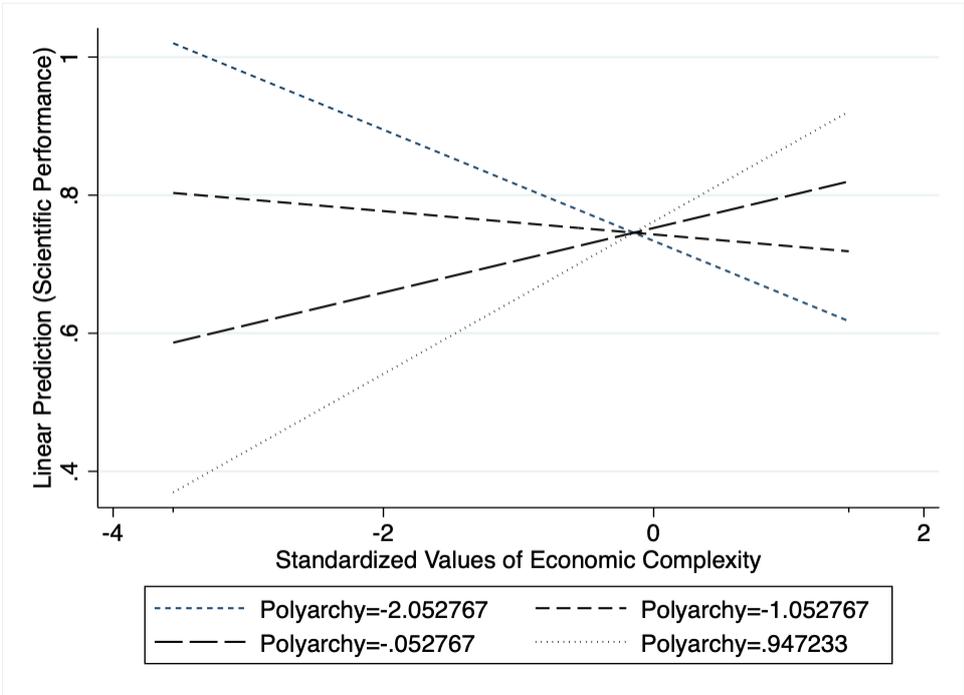

Figure Notes – The figure corresponds to model 2 in table 4. The figure shows the margin plot extracted from the fixed effect model

Figure 3 shows the margin plot for the interaction effect of polyarchy and globalization on Frac_FWCI. The figure shows that the higher level of polyarchy has a flat association on Frac_FWCI as globalization increases, while lower levels of polyarchy increase the association of globalization on Frac_FWCI. Contrary to the hypothesis, this indicates that the science systems of less democratic countries may be benefiting more from globalization. The non-intersecting lines indicate the interaction is non-significant.

**Figure 3 - Interaction Effect of Polyarchy and Globalization on Scientific Performance**

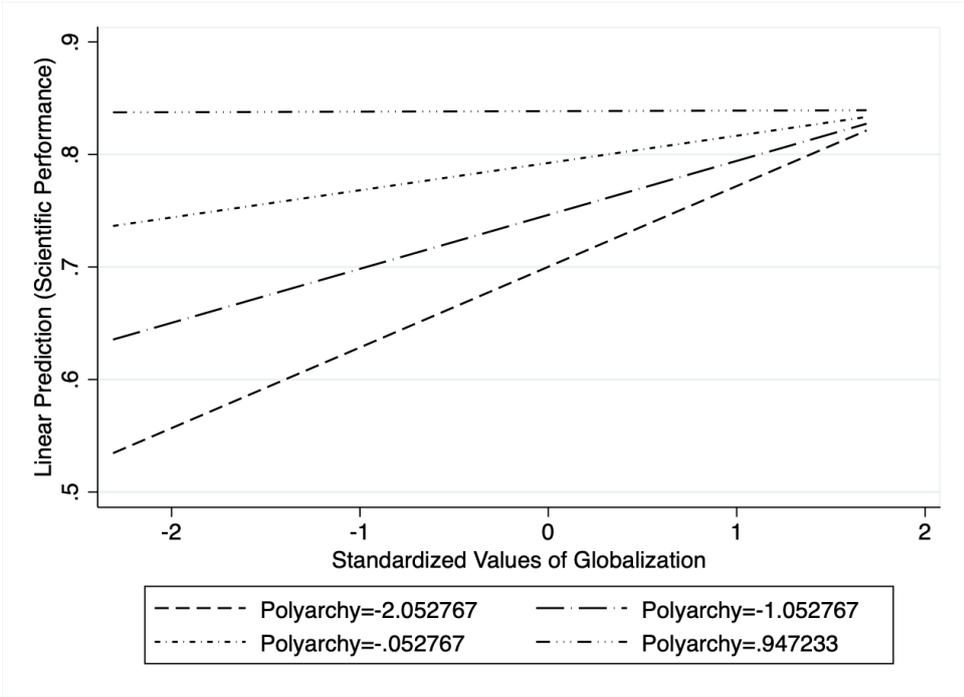

Figure Notes – The figure corresponds to model 4 in table 4. The figure shows the margin plot extracted from the fixed effect model.

Finally, Figure 4 shows the margin plot for the interaction effect of polyarchy and international collaboration on Frac_FWCI. The margin plot shows a similar interaction effect to Figure 2, where the highest level of polyarchy shows a flat line on Frac_FWCI as the values on international collaboration increase. As the level of polyarchy declines, the association with international collaboration increases. Contrary to the hypothesis, this suggests that less democratic countries may be benefiting more from international collaboration.

**Figure 4 - Interaction Effect of Polyarchy and Collaboration Intensity on Scientific Performance**

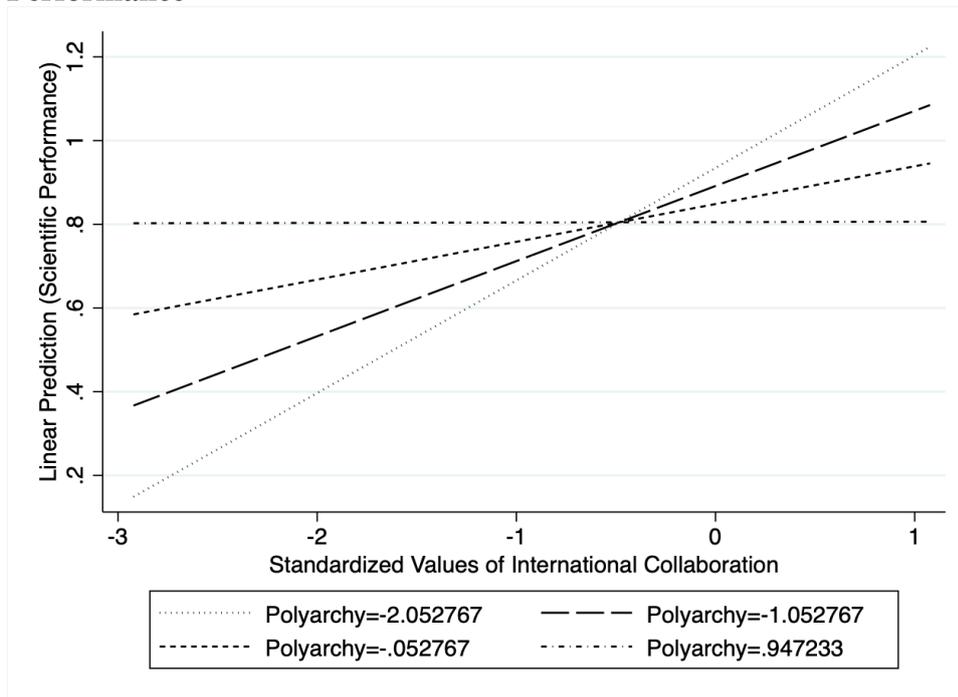

Figure Notes – The figure corresponds to model 6 in table 4. The figure shows the margin plot extracted from the fixed effect model

## 5. Discussion

We began this work seeking to test a long-standing research question regarding the relationship between democratic governance and national scientific performance. The results provided strong support for the subject line expectations. Specifically, the first hypothesis, which suggests that democratic governance is associated with scientific performance, was supported by the analysis, showing significant effects of all five high-level democracy indicators on scientific performance. Egalitarian democracy showed the strongest association. The results also support the second general research question regarding the moderating effects of complexity on the democracy-science relationship. Specifically, the second hypothesis, which suggests that economic complexity moderates the democracy-science relationship, was supported by the analysis, showing significant positive interaction effects between democracy and economic complexity on scientific performance. However, the analysis contradicted H3, which suggests that globalization moderates the democracy-science relationship, and H4, which suggests that collaboration intensity moderates the democracy-science relationship. The results showed significant negative interaction effects between democracy and globalization on scientific performance, and between democracy and international collaboration on scientific performance. Table 5 summarizes the implications of the research for the hypotheses.

**Table 5 – Summary of Results**

| Hypothesis | Dependent Variable | Key Independent Variable | Result |
|---|---|---|---|
| **H1:** Democratic governance is positively associated with the scientific performance of countries in the international system. | Scientific Performance | Five Democracy Indices | + |
| **H2:** Economic complexity positively moderates the democracy-science relationship among countries in the international system. | Scientific Performance | Polyarchy*Econ_Complexity | + |
| **H3:** Globalization positively moderates the democracy-science relationship among countries in the international system. | Scientific Performance | Polyarchy*Globalization | - |
| **H4:** International collaboration positively moderates the democracy-science relationship among countries in the international system. | Scientific Performance | Polyarchy*Collab_Intensity | - |

The results provide empirical support for a body of conceptual work done by philosophers and social science scholars over centuries. The question whether democratic governance is generically a positive environment for effective conduct of scientific inquiry has been one which has mostly been conjecture from armchair theorists and was frequently supported or challenged by analysis of cases for and against the hypothesis generated from the anecdotes of history. Many of these cases have given significant justification for believing that theory would not hold up in the face of broad empirical testing. More recently, Gao et al. (2017) provided a confounding study to the hypothesis, tested on the adjacent domain of technological innovation. Further, prominent cases seem to defy the hypothesis. For example, Russia, has become more democratic since the fall of the Soviet Union, yet its science system has suffered greatly, largely due to underinvestment and lack of state interest. China, on the other hand, has authoritarian governance yet has experienced gains in scientific performance. Yet the current study demonstrates that on average countries in the international system that experience increases or decreases in democracy also experience increases or decreases in scientific performance.

The study goes beyond the basic democracy-science compatibility thesis, recasting democratic governance itself as a type of complex system (e.g. Simon 1996), which permits greater internal and external heterogeneity and complexity than autocratic governance permits. As such the study advanced a second research question of, how does complexity moderate the democracy-science relationship? We expected to find that internal complexity has an upward moderating effect on the democracy-science effect. The results supported this expectation, showing that the internal knowledge intensity of nations, i.e. economic complexity, appears to increase the positive association between democracy and scientific performance. This result speaks to the theory that democracy itself may permit a greater level of structural complexity than other governance forms (e.g. Wiesner et al. 2018).

However, the study revealed something interesting about the interaction between democratic governance and external complexity, measured through globalization and international collaboration. We expected to find that external complexity moderates the democracy-science association upward. Yet the results show something less straightforward and perhaps more troubling. It appears that increases in scientific performance among advanced democracies do not hinge on globalization or international collaboration. Rather, the reverse seems to be correct. Autocratic and less democratic countries are experiencing increases in their levels of scientific performance at least in part because of globalization and international collaboration with the advanced democracies of the world. There appears to be directionality in the flow of knowledge. While historical evidence supports the claim that democracy provides fertile ground for the flourishing of science, countries with centralized forms of autocratic governance appear to have gained scientific capacity by collaborating with more decentralized democracies. Thus, advanced autocracies have been able to build scientific capacity without the emergent, self-organizing system that gave rise to the science in the first place. This raises a follow-on question, could these these less democratic countries have developed robust scientific systems without reliance on systems embedded in democratic governance structures. This question remains unanswered and perhaps unanswerable. It is unclear how long this trend might last, whether there is a point at which globalization and collaboration can only go so far, and whether/what proportion of the result is a function of the mere fact that less democratic countries simply have more room for growth in the global network of science.

**5.1.** *Limitations and Future Research*: As with all non-experimental data, the study cannot rule out threats to internal validity, such as the influence of unique historical conditions or selection threats. One concern that readers may have is whether high performing countries like the USA may be skewing the results. To address this concern, we reran the bivariate correlations and the fixed effects models dropping the USA, finding no meaningful differences. Another concern is that our data on democracy and science come from a highly developed period in the 21st century when many countries had already established systems of governance that only changed in small ways. The concern of relatively low variation on the key independent variable means that larger stretches of time will be useful in future testing. While there was sufficient variation within the 124 countries over the eleven-year period for the fixed effects models, more years of data would strengthen the external validity of the models. Another limitation of the study was uneven data coverage over all the variables, e.g. the moderation models had a lower sample size due to lack of data coverage on many countries. Further, measures of science that rely on bibliometric statistics alone may have limitations in covering scientific advancements that may be more qualitative or less straightforwardly captured in the institutions of science. Finally, the prospect of reverse causality also poses an interesting question, to what extent does advancement in science effect change in governance systems? For these reasons, we avoid causal claims based on the results and suggest rather that the democracy-science hypothesis has yet to be falsified.

We identify potential avenues for future research. First, the sub-dimensions of democracy and scientific performance could be further explored to estimate any specific impacts of more nuanced elements of democracy in sub-domains of science. This will require more theory development to produce plausible hypotheses. Second, alternative indicators for internal and external complexity could also be developed and tested, including those for economic complexity, as well as alternative indicators for national scientific performance. Third, as the data are updated annually, it will be interesting to observe how the results change over time commensurate with changes in global conditions. Finally, the dynamic relationship between democracy, science, and external complexity could be further explored to describe advancements in science in developing and less democratic countries. Pertinent questions remain, such as how much of the scientific advancement among non-democratic nations is due to collaboration with developed democracies, what is the effect of such collaboration on democracy and vis-versa, and what are the sustainability concerns in the long term given recent anti-democratic trends globally?

## 6. Conclusion

Until very recently, the thesis that democratic governance is requisite for scientific advancement has mostly been asserted by Western philosophers through the examination of numerous cases during the Cold War Era. It has been difficult to test, yet recent advancements in bibliometric techniques used to track the development and influence of scientific activity have provided new means for establishing the relative advantage of countries in science. This study puts the democracy-science compatibility thesis to the test by analysing the relationship between measures of democratic governance and scientific performance among a large sample of 124 countries over eleven years in the early 21$^{st}$ century. This study shows significant estimates of democracy on scientific performance, using fixed effects panel data analysis, controlling for number of scientific papers, GDP per capita, and population size. Finally, this research finds interesting interaction effects of internal and external complexity, where internal economic complexity enhances the democracy-science relationship, and interactions with globalization and international collaboration appear to have the opposite effect. This surprising result potentially indicates that less democratic countries in the international system are reaping gains in scientific performance through increases in external complexity.

## 8. Appendix. Distribution Analysis and Robustness Checks

Below we show the frequency distributions for Frac_FWCI and polyarchy both before and after removing outliers from the distributions. As the distributions show Frac_FWCI is sensitive to outliers from very low-capacity countries that almost exclusively collaborate with high impact developed countries. As the distribution analysis shows, filtering out countries below the median for total publications proved useful in normalizing the distribution.

Figure A.1 - Frequency Distribution Comparison with/without outliers.

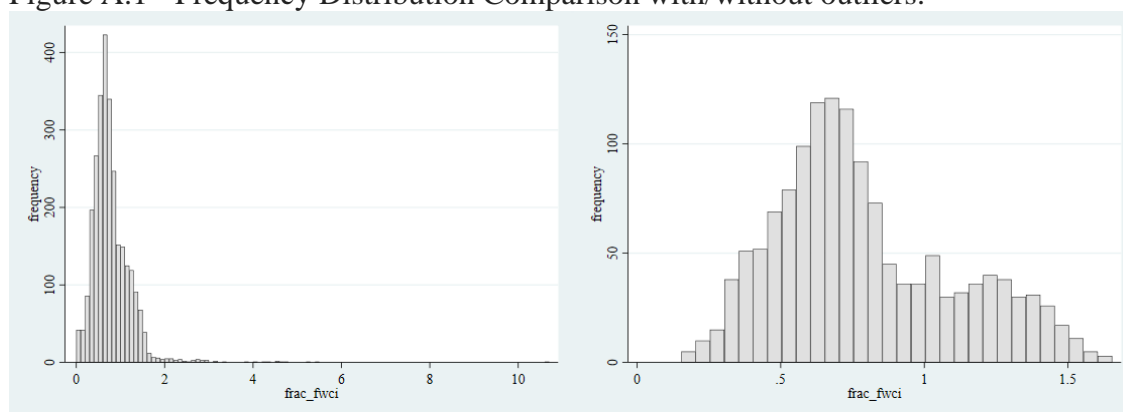

Next, we show the GMM robustness checks. Dynamic panel models are subject to "three main sources for correlation in the dependent variable over time: true state dependence, due to the inclusion of preceding periods; observed heterogeneity directly through the explanatory variables; and unobserved heterogeneity – indirectly through time-invariant or fixed effects" (Zuazu-Bermejo, 2015, p 11; Cameron & Trivedi, 2005). GMM models overcome these types of correlations. "System" GMM, as opposed to "difference" GMM, relies not only on first-differenced data for eliminating fixed effects but estimates a system of two equations – in both difference and in levels, the original and the transformed one (Roodman, 2009). Because at times it is difficult or impossible to identify the perfect instruments, we draw from instruments readily available in our dataset (Roodman, 2009). We transform the models from tables 3 and 4 into dynamic panel data models by including the lagged dependent variable (scientific impact lagged) into the set of independent variables (please refer to tables A.1 and A.2). However, GMM may be subject to instrument proliferation (Roodman, 2009). To limit the number of instruments, we "collapse" the internal temporal lags following Roodman (2009)'s recommendation. Overall the GMM regression results are consistent with the fixed effects models. Regression outputs are available upon request.

**Table A.1. Effects of Democracy on Scientific Performance (GMM Estimators)**

|  | Model 1 | Model 2 | Model 3 | Model 4 | Model 5 |
|---|---|---|---|---|---|
| **L. Scientific Performance** | 0.613*** | 0.611*** | 0.604*** | 0.626*** | 0.599*** |
|  | (0.07) | (0.07) | (0.07) | (0.07) | (0.07) |
| **L. Scientific Performance** | 0.196** | 0.195** | 0.201*** | 0.217*** | 0.205*** |
|  | (0.06) | (0.06) | (0.06) | (0.06) | (0.06) |
| **Polyarchy** | 0.183* |  |  |  |  |
|  | (0.09) |  |  |  |  |
| **Liberal_Democracy** |  | 0.193* |  |  |  |
|  |  | (0.09) |  |  |  |
| **Participatory_Democracy** |  |  | 0.232 |  |  |
|  |  |  | (0.12) |  |  |
| **Deliberative_Democracy** |  |  |  | 0.152* |  |
|  |  |  |  | (0.07) |  |
| **Egalitarian_Democracy** |  |  |  |  | 0.204* |
|  |  |  |  |  | (0.10) |
| **Tot_Publications** | 0.000 | 0.000 | 0.000 | 0.000 | 0.000 |
|  | (0.00) | (0.00) | (0.00) | (0.00) | (0.00) |
| **GDP_Per_Capita** | 0.000 | 0.000 | 0.000 | 0.000 | 0.000 |
|  | (0.00) | (0.00) | (0.00) | (0.00) | (0.00) |
| **Population_Size** | 0.000 | 0.000 | 0.000 | -0.000 | 0.000 |
|  | (0.00) | (0.00) | (0.00) | (0.00) | (0.00) |
| N | 993 | 993 | 993 | 993 | 993 |
| #Countries | 124 | 124 | 124 | 124 | 124 |
| AR1 | 0.001 | 0.002 | 0.002 | 0.002 | 0.002 |
| AR2 | 0.829 | 0.841 | 0.730 | 0.721 | 0.752 |
| Hansen p-value | 0.220 | 0.201 | 0.181 | 0.186 | 0.202 |
| Diff-in-Hansen (excluding group) | 0.175 | 0.208 | 0.125 | 0.160 | 0.211 |
| Diff-in-Hansen (H0=exogenous) | 0.591 | 0.336 | 0.709 | 0.489 | 0.330 |
| Number of instruments | 55 | 55 | 55 | 55 | 55 |

Notes
AR rows report the p-value for the test of serial correlation in the residuals.
Hansen test for overidentification restrictions and exogeneity of instruments.
Standard errors in parentheses; * p<0.05, **, p<0.01, *** p<0.001

**Table A.2. – Moderation Effects of Complexity (GMM Estimators)**

|  | Model 1 | Model 2 | Model 3 | Model 4 | Model 5 | Model 6 | Model 7 |
|---|---|---|---|---|---|---|---|
| L. Scientific Performance | 0.812*** | 0.719*** | 0.659*** | 0.641**** | 0.609*** | 0.462*** | 0.582*** |
| | (0.07) | (0.07) | (0.07) | (0.06) | (0.07) | (0.07) | (0.06) |
| L. Scientific Performance | 0.215* | 0.205* | 0.238*** | 0.243*** | 0.256*** | 0.253*** | 0.306*** |
| | (0.11) | (0.08) | (0.06) | (0.05) | (0.06) | (0.05) | (0.07) |
| Polyarchy | 0.023 | 0.000 | 0.050* | 0.049 | 0.039 | -0.011 | -0.047 |
| | (0.03) | (0.02) | (0.02) | (0.03) | (0.02) | (0.02) | (0.04) |
| Econ_Complexity | -0.003 | 0.021 | | | | | 0.021 |
| | (0.02) | (0.02) | | | | | (0.02) |
| Polyarchy X Econ_Complexity | | 0.045** | | | | | -0.002 |
| | | (0.02) | | | | | (0.03) |
| Globalization_Index | | | 0.011 | 0.024 | | | -0.019 |
| | | | (0.03) | (0.03) | | | (0.04) |
| Polyarchy X Globalization_Index | | | | -0.016 | | | 0.024 |
| | | | | (0.03) | | | (0.03) |
| Collab_Intensity | | | | | 0.050* | 0.060** | 0.046 |
| | | | | | (3.03) | (0.02) | (0.03) |
| Polyarchy X Collab_Intensity | | | | | | -0.070*** | -0.046 |
| | | | | | | (0.02) | (0.02) |
| Tot_Publications | 0.000 | 0.000 | 0.000 | 0.000 | 0.000* | 0.000*** | 0.000* |
| | (0.00) | (0.00) | (0.00) | (0.00) | (0.00) | (0.00) | (0.00) |
| GDP_Per_Capita | -0.000 | -0.000 | -0.000 | -0.000 | 0.000 | 0.000 | 0.000 |
| | (0.00) | (0.00) | (0.00) | (0.00) | (0.00) | (0.00) | (0.00) |
| Population_Size | -0.000 | -0.000 | 0.000 | 0.000 | 0.000 | -0.000 | -0.000 |
| | (0.00) | (0.00) | (0.00) | (0.00) | (0.00) | (0.00) | (0.00) |
| N | 650 | 650 | 974 | 974 | 993 | 993 | 643 |
| #Countries | 100 | 100 | 121 | 121 | 124 | 124 | 99 |
| AR1 | 0.002 | 0.001 | 0.002 | 0.002 | 0.003 | 0.004 | 0.002 |
| AR2 | 0.793 | 0.846 | 0.553 | 0.487 | 0.332 | 0.167 | 0.499 |
| Hansen p-value | 0.124 | 0.283 | 0.274 | 0.238 | 0.007 | 0.012 | 0.369 |
| Diff-in-Hansen (excluding group) | 0.171 | 0.284 | 0.329 | 0.260 | 0.013 | 0.015 | 0.229 |
| Diff-in-Hansen (H0=exogenous) | 0.181 | 0.385 | 0.235 | 0.311 | 0.105 | 0.205 | 0.847 |
| # of Instruments | 54 | 63 | 66 | 77 | 66 | 77 | 99 |

Table notes:
AR rows report the p-value for the test of serial correlation in the residuals.
Hansen test for overidentification restrictions and exogeneity of instruments.
N varies across models due to uneven data coverage of variables
Standard errors in parentheses; * p<0.05, **, p<0.01, *** p<0.001

We also conducted additional robustness checks using Bayesian mixed effects panel data regression, using the bayes prefix in the mixed procedure in Stata. A thinning factor of 3 was used to avoid autocorrelation. Default normal priors were used. The results are below in table A.3 and A4. The models derive point estimates by simulating a posterior distribution using Markov chain Monte Carlo estimation on each parameter. The significance of the mean estimates are interpreted using the credibility intervals, which are shown in parentheses. The credibility estimates should not contain zero, and the spread in the range indicates the accuracy of the estimate.

**Table A.3. Effects of Democracy on Scientific Performance (Bayes Mixed Effects Regression)**

|  | Model 1 | Model 2 | Model 3 | Model 4 | Model 5 |
|---|---|---|---|---|---|
| Polyarchy | 0.22 | | | | |
|  | (0.14, 0.29) | | | | |
| Liberal_Democracy | | 0.22 | | | |
|  | | (0.14, 0.30) | | | |
| Participatory_Democracy | | | 0.30 | | |
|  | | | (0.19, 0.40) | | |
| Deliberative_Democracy | | | | 0.19 | |
|  | | | | (0.12, 0.26) | |
| Egalitarian_Democracy | | | | | 0.29 |
|  | | | | | (0.20 0.39) |
| Tot_Publications | 1.13e-06 | 1.10e-06 | 1.15e-06 | 1.20e-06 | 1.14e-06 |
|  | (6.56e-07, 1.58e-06) | (6.65e-07, 1.59e-06) | (6.55e-07, 1.65e-06) | (7.13e-07, 1.68e-06) | (6.66e-07, 1.62e-06) |
| GDP_Per_Capita | 5.72e-06 | 5.81e-06 | 5.88e-06 | 5.78e-06 | 5.65e-06 |
|  | (4.57e-06, 6.90e-06) | (4.56e-06, 7.02e-06) | (4.67e-06, 7.04e-06) | (4.60e-06, 6.89e-06) | (4.54e-06, 6.84e-06) |
| Population_Size | -1.17e-10 | -1.04e-10 | -1.06e-10 | -1.36e-10 | -1.05e-10 |
|  | (-3.78e-10, 1.61e-10) | (-3.48e-10, 1.51e-10) | (-3.30e-10, 1.34e-10) | (-3.75e-10, 1.33e-10) | (-3.36e-10,1.39e-10) |
| MCMC iterations | 32,498 | 32,498 | 32,498 | 32,498 | 32,498 |
| Burn-in | 2,500 | 2,500 | 2,500 | 2,500 | 2,500 |
| Obs | 1,188 | 1,188 | 1.188 | 1.188 | 1.188 |
| Acceptance rate | 0.81 | 0.81 | 0.81 | 0.80 | 0.81 |
| Efficiency (min) | 0.006 | 0.007 | 0.009 | 0.004 | 0.007 |

Note: We report the mean value and L95% L.CI and 95% U.CI in parenthesis below (L.CI=lower credible interval; U.CI = upper credible interval). Scientific notation used for very small numbers

**Table A.4. – Moderation Effects of Complexity (Bayes Mixed Effects Regression)**

|  | Model 1 | Model 2 | Model 3 | Model 4 | Model 5 | Model 6 | Model 7 |
|---|---|---|---|---|---|---|---|
| Polyarchy | 0.04 | 0.026 | 0.04 | 0.04 | 0.06 | -0.02 | -0.03 |
|  | (0.22, 0.06) | (0.004, 0.05) | (0.02, 0.06) | (0.02, 0.07) | (0.04, 0.07) | (-0.04, 0.006) | (-0.05, – 0.0002) |
| Econ_Complexity | 0.02 | 0.03 |  |  |  |  | 0.04 |
|  | (-0.00, 0.05) | (0.003, 0.05) |  |  |  |  | (0.02, 0.06) |
| Polyarchy X Econ_Complexity |  | 0.06 |  |  |  |  | 0.05 |
|  |  | (0.03, 0.08) |  |  |  |  | (0.02, 0.07) |
| Globalization_Index |  |  | 0.08 | 0.08 |  |  | 0.07 |
|  |  |  | (0.05, 0.12) | (0.04, 0.11) |  |  | (0.03, 0.11) |
| Polyarchy X Globalization_Index |  |  |  | -0.00 |  |  | -0.03 |
|  |  |  |  | (-0.03, 0.02) |  |  | (-0.06, -0.005) |
| Collab_Intensity |  |  |  |  | 0.11 | 0.09 | 0.08 |
|  |  |  |  |  | (0.09, 0.13) | (0.07, 0.11) | (0.05, 0.10) |
| Polyarchy X Collab_Intensity |  |  |  |  |  | -0.08 | -0.06 |
|  |  |  |  |  |  | (-0.10, – 0.07) | (-0.08, -0.05) |
| Tot_Publications | 7.57e-07 | 8.03e-07 | 1.03e-06 | 1.06e-06 | 1.23e-06 | 1.16e-06 | 7.50e-07 |
|  | (3.19e-07, 1.17e-06) | (7.99e-07, 3.51e-07) | (5.39e-07, 1.52e-06) | (5.66e-07, 1.55e-06) | (7.82e-07, 1.73e-06) | (7.48e-07, 1.62e-06) | (3.50e-07, 1.15e-06) |
| GDP_Per_Capita | 6.79e-06 | 7.12e-06 | 4.69e-06 | 4.64e-06 | 4.00e-06 | 5.03e-06 | 4.90e-06 |
|  | (5.48e-06, 8.03e-06) | (7.18e-06, 5.65e-06) | (3.45e-06, 5.95e-06) | (3.40e-06, 5.94e-06) | (2.91e-06, 5.11e-06) | (3.85e-06, 6.15e-06) | (3.68e-06, 6.18e-06) |
| Population_Size | -1.22e-10 | -1.18e-10 | -8.93e-11 | -1.22e-10 | -7.55e-11 | 1.52e-13 | -7.59e-11 |
|  | (-3.71e-10, 9.69e-11) | (-4.04e-10, 1.44e-10) | (-3.28e-10, 1.48e-10) | (-3.53e-10, 1.24e-10) | (-3.39e-10, 1.66e-10) | (-2.43e-10, 2.40e-10) | (-3.44e-10, 1.80e-10) |
| MCMC iterations | 32,489 | 32,489 | 32,489 | 32,489 | 32,489 | 32,489 | 32,489 |
| Burn-in | 2,500 | 2,500 | 2,500 | 2,500 | 2,500 | 2,500 | 2,500 |
| Obs | 831 | 831 | 1,167 | 1,167 | 1,188 | 1,188 | 822 |
| Acceptance rate | 0.81 | 0.81 | 0.81 | 0.81 | 0.81 | 0.81 | 0.81 |
| Efficiency (min) | 0.004 | 0.004 | 0.005 | 0.005 | 0.007 | 0.004 | 0.003 |

Note: We report the mean value and L95% L.CI and 95% U.CI in parenthesis below (L.CI=lower credible interval; U.CI = upper credible interval). Scientific notation used for very small numbers